\documentclass[letterpaper]{article}
\usepackage[preprint]{aaai2027}
\usepackage[hyphens]{url}
\usepackage{graphicx}
\usepackage{natbib}
\usepackage{caption}
\usepackage{booktabs}
\usepackage{amsmath,amssymb}
\usepackage{tcolorbox}
\newcommand{\method}{LineageRAG}
\newtcolorbox{promptbox}[1]{
  colback=white,
  colframe=black,
  colbacktitle=white,
  coltitle=black,
  coltext=black,
  fonttitle=\bfseries\normalsize,
  fontupper=\normalsize,
  title={#1},
  boxrule=0.4pt,
  titlerule=0pt,
  arc=0pt,
  boxsep=0pt,
  left=4pt,
  right=4pt,
  top=1pt,
  bottom=3pt,
  lefttitle=4pt,
  righttitle=4pt,
  toptitle=2pt,
  bottomtitle=1pt,
  before skip=3pt,
  after skip=3pt
}

\title{LineageRAG: Harnessing GraphRAG by Constructing Evidence Lineages\\
with Source Grounding}
\author{
Linyao Zheng\textsuperscript{\rm 2}\thanks{Corresponding author. Email: \texttt{zlyxjtu@gmail.com}},
Xuhang Shi\textsuperscript{\rm 1},
Zhifang Mao\textsuperscript{\rm 1},\\
Sai Zhou\textsuperscript{\rm 2},
Shuaixian An\textsuperscript{\rm 2},
Xiuquan Hou\textsuperscript{\rm 2},
Jinze Li\textsuperscript{\rm 2}
}
\affiliations{
\textsuperscript{\rm 1}XiaoLab, Beijing University of Posts and Telecommunications
\quad
\textsuperscript{\rm 2}Xi'an Jiaotong University
}

\begin{document}
\maketitle
\enlargethispage{-22pt}

\begin{abstract}
Graph-based Retrieval-Augmented Generation (GraphRAG) retrieves evidence for
multi-hop questions over structured corpus graphs. Existing GraphRAG methods
leave the connection between evidence discovery and source grounding implicit.
We propose LineageRAG, which constructs one evidence lineage for each
query-derived evidence demand and completes it with a verbatim source span when
the selected evidence supports that demand. LineageRAG first initializes the
evidence demands. It then expands each lineage through demand-conditioned
retrieval over the corpus graph while retaining the demand associated with
every candidate. Lineage completion uses this provenance to select
complementary passages and grounds supported demands in verbatim source text.
Experiments on HotpotQA, 2WikiMultiHopQA, and MuSiQue show that LineageRAG
improves R@5, EM, and F1 by 3.51, 5.96, and 5.22 points on average over
leading GraphRAG baselines.
\end{abstract}

\section{Introduction}

\begin{figure*}[t]
  \centering
  \includegraphics[width=\textwidth]{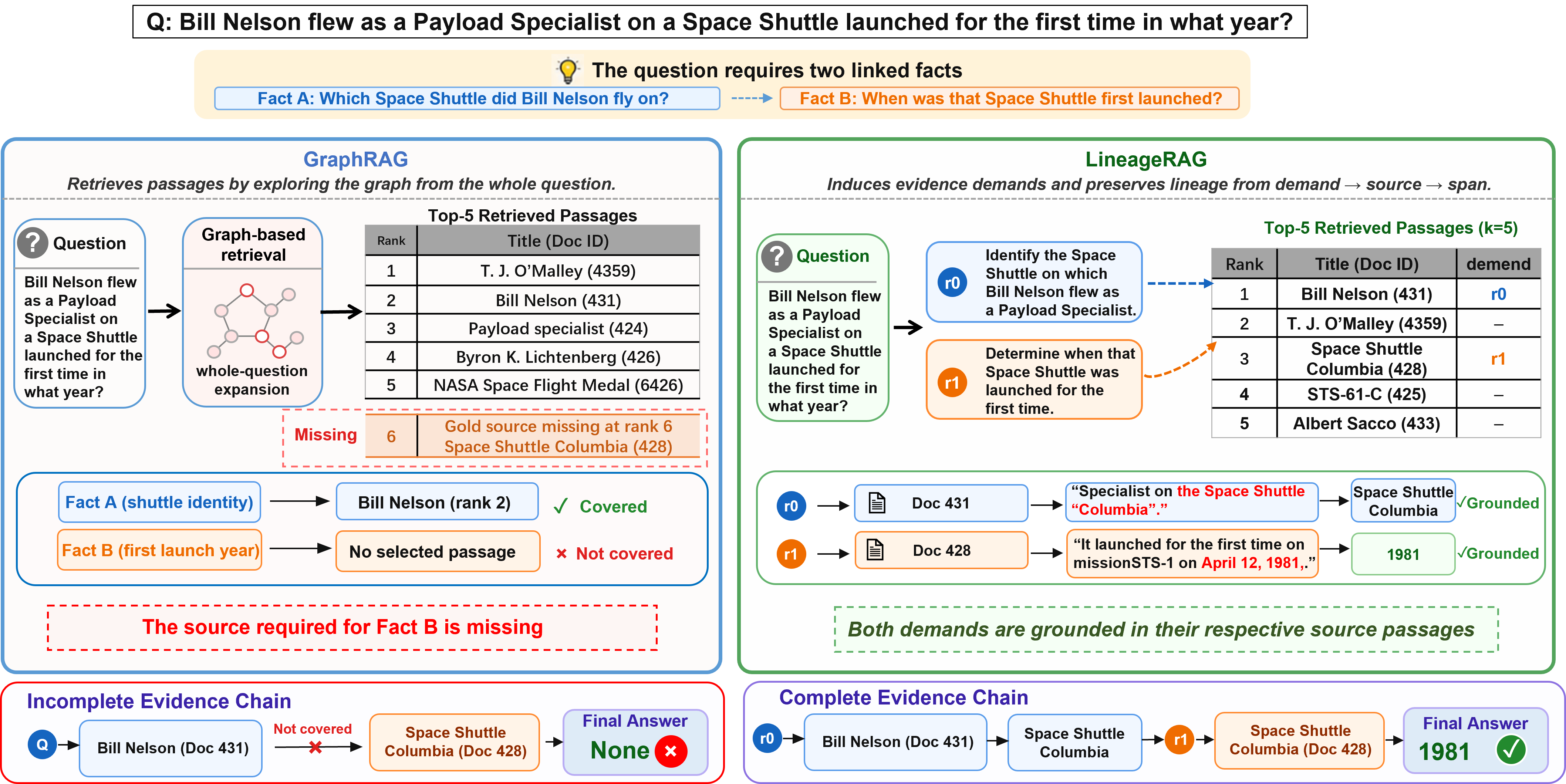}
  \caption{A GraphRAG failure case on HotpotQA under R@5. Whole-question
  ranking misses the Columbia passage required for the second hop and leaves
  the evidence chain incomplete. \method{} retrieves evidence for both hops
  using separate demands and returns 1981 as the correct answer.}
  \label{fig:motivation}
\end{figure*}

Retrieval-augmented generation (RAG) supplies a language model with evidence
retrieved from an external corpus \citep{lewis2020rag}. Conventional retrievers
score passages independently, which works when one passage contains the required
evidence. Multi-hop questions often depend on bridge passages whose relevance
emerges through previously discovered evidence. Graph-based RAG (GraphRAG)
addresses this dependency by organizing corpus units through structured
relations. HippoRAG propagates relevance across an OpenIE graph using
Personalized PageRank \citep{gutierrez2024hipporag}. HippoRAG~2 deepens passage
integration through online language-model use \citep{gutierrez2025hipporag2}.
PropRAG retrieves proposition paths \citep{wang2025proprag}. HGRAG uses a
cross-granularity hypergraph \citep{wang2026hgrag}. Graph-based retrieval finds
relevant candidates, but the final evidence set does not record the requirement
behind each source.
This missing connection can leave GraphRAG with an incomplete evidence chain.
Figure~\ref{fig:motivation} presents a representative GraphRAG failure on
HotpotQA. The retrieved passages cover only part of the question, leaving one
evidence demand unresolved. This example motivates our central question.
\textbf{\emph{How can GraphRAG carry each evidence demand from graph discovery
to source grounding?}}

Previous work uses query structure at different points in retrieval. Question
decomposition preserves subqueries during candidate discovery
\citep{ammann2025decomposition}. SETR uses explicit information requirements
during evidence selection \citep{lee2025setr}, while RichRAG retains
aspect--document associations during ranking \citep{wang2025richrag}. Relink
organizes retrieved evidence into a query-specific graph
\citep{huang2026relink}. NeocorRAG organizes retrieval around evidence chains
\citep{peng2026neocorrag}. These methods improve coverage during
discovery, ranking, or selection, but do not carry the same demand--passage
association through graph retrieval to verbatim source evidence.

Lineage tracing offers a natural model for this missing history. In
biology, it identifies the descendants of a marked cell
\citep{kretzschmar2012lineage}. Data lineage applies the same principle to
derived records by identifying the source items that produced them
\citep{cui2000lineage}. For GraphRAG, we define an \textbf{\emph{evidence lineage}} as
the trace that carries one evidence demand from graph discovery to its verbatim
source evidence.

We propose \method{}, a GraphRAG framework that constructs and maintains
these lineages in three stages. \textbf{\emph{Evidence-Demand Induction}} derives
query-specific demands, each specifying what the retrieved evidence must
establish. \textbf{\emph{Demand-Conditioned Lineage Expansion}} retrieves candidates for
each demand and preserves their demand associations during candidate merging.
\textbf{\emph{Lineage Completion with Source Grounding}} uses this provenance
to select complementary passages and extract source-anchored verbatim spans from the
selected passages. Retaining such a span completes the corresponding lineage.

Our contributions are as follows.
\begin{itemize}
  \item We formulate evidence lineage as a trace that follows each
  Evidence Demand from graph discovery to verbatim source evidence. It keeps
  the provenance of each demand explicit while the final passage set is
  constructed.
  \item We introduce \method{}, which preserves each Evidence Demand's identity
  from candidate discovery to grounded source evidence. This correspondence
  guides passage selection and links each supported demand to a source-anchored
  span.
  \item Experiments on multi-hop QA benchmarks show that \method{} achieves the
  strongest overall results, with average gains of 3.51 R@5, 5.96 EM, and
  5.22 F1 points over leading GraphRAG baselines.
\end{itemize}

\section{Related Work}

\paragraph{GraphRAG.}
GraphRAG represents corpus knowledge or query-specific evidence as a graph and
retrieves through its relations. HippoRAG propagates relevance over an OpenIE
graph \citep{gutierrez2024hipporag}, and HippoRAG~2 strengthens the integration
of passages and graph retrieval \citep{gutierrez2025hipporag2}. PropRAG searches
proposition paths with an online beam search \citep{wang2025proprag}. LightRAG
combines graph and vector retrieval at two granularities
\citep{guo2025lightrag}, while HGRAG retrieves over a cross-granularity
hypergraph \citep{wang2026hgrag}. G-reasoner trains a graph foundation model
over its QuadGraph abstraction \citep{luo2026greasoner}. SUBQRAG stores triples
from sub-question reasoning as graph memory \citep{li2025subqrag}. AGRAG presents
a reasoning subgraph with the retrieved chunks \citep{wang2025agrag}. These
systems improve evidence discovery through graph structure. \method{} also keeps
the connection between each Evidence Demand and its retrieved sources available
during passage selection and source grounding.

\paragraph{Query decomposition and evidence selection.}
Question decomposition preserves subqueries during candidate discovery and
merges their results for reranking \citep{ammann2025decomposition}. DeCoR uses
decomposed queries and compressed evidence to initiate a second retrieval stage
\citep{yun2025decor}. SETR uses explicit information requirements during
evidence selection \citep{lee2025setr}, while RichRAG retains aspect--document
associations during ranking \citep{wang2025richrag}. Relink constructs a
query-driven evidence graph \citep{huang2026relink}, NeocorRAG organizes
retrieval around evidence chains \citep{peng2026neocorrag}, and ETS searches
over candidate evidence sets with a tree \citep{sun2025ets}. \method{} preserves
each Evidence Demand's identity through retrieval, selection, and grounding to a verbatim
source span.

\begin{figure*}[t]
  \centering
  \includegraphics[width=\textwidth]{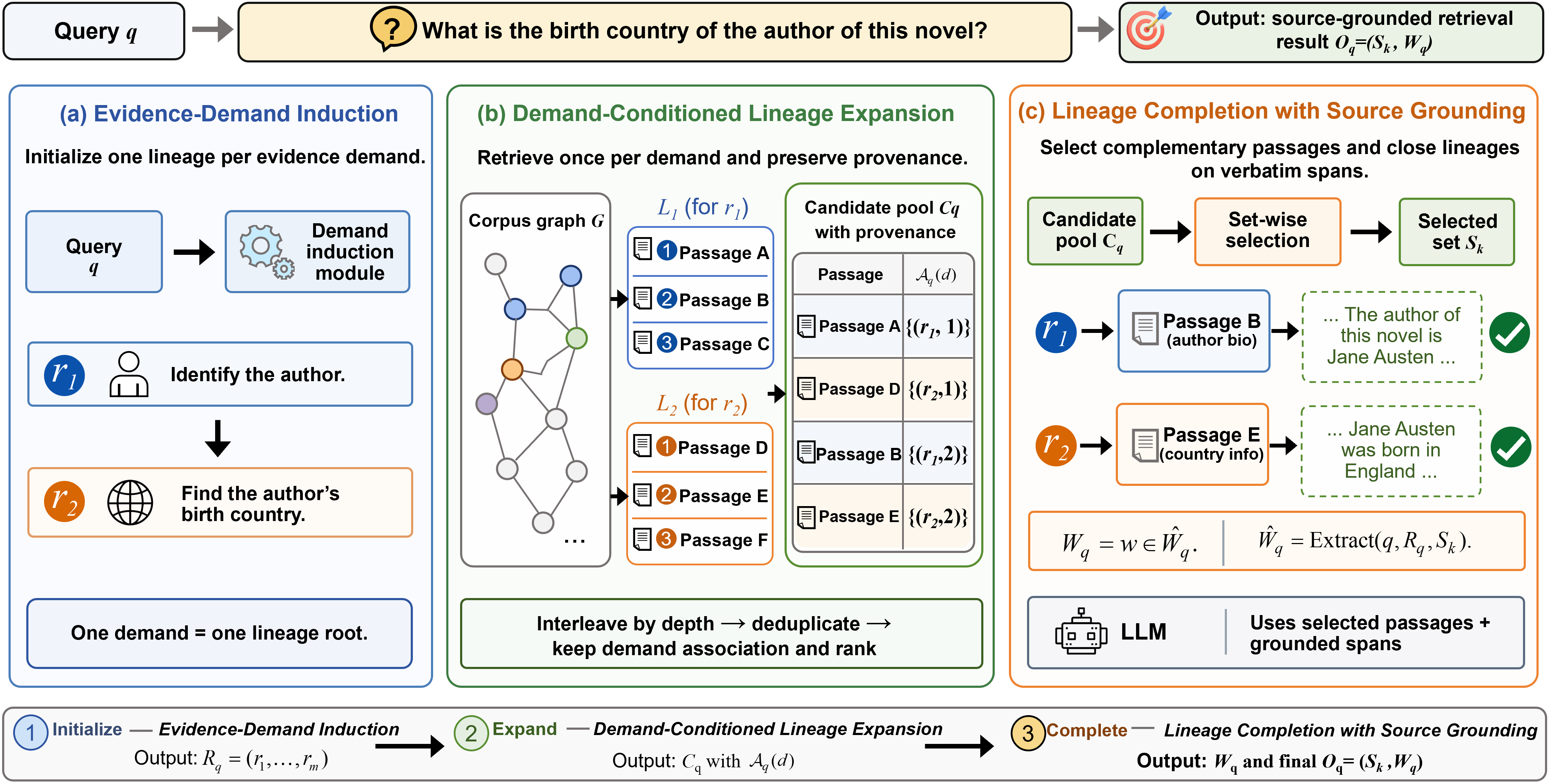}
  \caption{Overview of \method{}. Evidence-Demand Induction identifies facts
  required by the query. Demand-Conditioned Lineage Expansion retrieves candidates
  for each demand and preserves demand--passage provenance. Lineage Completion
  with Source Grounding selects complementary passages and grounds each supported
  demand in a verbatim source span.}
  \label{fig:method-overview}
\end{figure*}

\section{Problem Formulation}

Let $q$ be a query and $\mathcal D=\{d_i\}_{i=1}^{N}$ a passage corpus. For
$d\in\mathcal D$, $\operatorname{text}(d)$ denotes its source text.
$\mathcal G$ is the graph index over this corpus. The parameter $k$ specifies the context
size. An \emph{Evidence Demand} is a query-specific evidence requirement.
The method produces
\begin{equation}
\mathcal O_q=(S_k,W_q),\qquad S_k=(d^{(1)},\ldots,d^{(k)})\in\mathcal D^k,
\end{equation}
where $S_k$ is the ordered list of selected passages and $W_q$ is the set of
grounded annotations. The passages in $S_k$ are distinct. Each annotation
$w=(i_w,d_w,s_w,a_w)$ links demand $i_w$ to span $s_w$ in selected passage
$d_w$. The term $a_w$ records the resulting typed relation.

\section{LineageRAG}

\method{} carries each Evidence Demand from graph discovery to a grounded
source span in three stages. Figure~\ref{fig:method-overview} illustrates these
stages with an example.

\subsection{Evidence-Demand Induction}

An inference-time language model maps $q$ to an ordered sequence of Evidence
Demands using a fixed prompt $\mathcal P_D$:
\begin{equation}
R_q=\operatorname{Normalize}(\operatorname{LLM}_{\theta}(\mathcal P_D,q))=(r_1,\ldots,r_m).
\end{equation}
\begin{equation}
r_j=(\mathrm{id}_j,\delta_j,I_j,O_j),\qquad 1\leq j\leq m\leq5.
\end{equation}
The description $\delta_j$ states the fact that evidence must establish. $I_j$ records the
symbolic variables consumed by the demand, and $O_j$ records those its evidence
should establish. They are not values resolved from retrieved evidence. All
demands are induced jointly from $q$. In Figure~\ref{fig:motivation}, $r_0$
identifies the shuttle and $r_1$ requests its launch year. The identifier
$\mathrm{id}_j$ remains fixed through retrieval, selection, and grounding.

The induction prompt asks the model to express the required facts as ordered
demands:
\begin{promptbox}{Evidence-Demand Induction Prompt $\mathcal P_D$}
Decompose the query into ordered Evidence Demands whose joint resolution makes
the answer derivable. Each demand states one fact that a source passage must
establish. Separate evidence identifying an intermediate entity from evidence
establishing the answer.
\end{promptbox}
The complete prompt and JSON schema are provided in the supplementary material.

\subsection{Demand-Conditioned Lineage Expansion}

For each Evidence Demand, \method{} retrieves a ranked passage list from the
same graph index:
\begin{equation}
L_j=\operatorname{Retrieve}_{\mathcal G}(q,\delta_j)
=(d_{j,1},\ldots,d_{j,h}),
\end{equation}
where $h$ is the retrieval depth. The retrieval query formed from $q$ and
$\delta_j$ determines the graph seeds. Personalized PageRank propagates the seed
scores over $\mathcal G$ and ranks passages to form $L_j$. Each call uses
$(q,\delta_j)$ independently and does not substitute values resolved for earlier
demands.

The order of $R_q$ controls depth-wise list merging and demand presentation to
selection and grounding. A passage enters $C_q=(c_1,\ldots,c_M)$ at its first
occurrence, with $M\geq k$. Let
$\tau(d)$ be its normalized title. The retained provenance and selector
representation are
\begin{equation}
\mathcal A_q(d)=\{(\mathrm{id}_j,\ell)\mid d=d_{j,\ell}\}.
\end{equation}
\begin{equation}
b_q(d)=(\mathcal A_q(d),n_q(d),\rho_q(d),g_q(d)),\qquad d\in C_q.
\end{equation}
$\mathcal A_q(d)$ preserves the demand ID and rank of every retrieval event that
exposed $d$. The count $n_q(d)=|\mathcal A_q(d)|$ summarizes repeated exposure.
The statistic $\rho_q(d)$ is the best rank recorded in $\mathcal A_q(d)$.
Same-title redundancy is represented by
$g_q(d)=|\{c\in C_q:\tau(c)=\tau(d)\}|$.

During merging, \method{} retains each candidate's demand IDs and within-demand
ranks in $\mathcal A_q(d)$. The selector can then distinguish candidates
retrieved for different Evidence Demands and construct a passage set that covers
complementary requirements.

\subsection{Lineage Completion with Source Grounding}

\paragraph{Set-wise selection.}
The selector evaluates the merged candidate pool against $R_q$. Each candidate
excerpt is augmented by $b_q(d)$:
\begin{equation}
S_k=\operatorname{Select}_k
(q,R_q,C_q,\{b_q(d)\}_{d\in C_q}).
\end{equation}
The selector evaluates candidates jointly to build the passage set. The provenance in
$b_q(d)$ reveals whether a candidate contributes to an
uncovered demand. It can therefore favor a complementary passage even when a
redundant passage is ranked higher in the merged pool.

The selector applies this provenance through the following instruction:
\begin{promptbox}{Set-Wise Selection Prompt}
Use each candidate's demand-specific retrieval provenance to select $k$
passages that jointly cover the Evidence Demands. Favor complementary facts and
avoid evidence that repeats a covered demand.
\end{promptbox}
Returned IDs are validated against $C_q$ and deduplicated. The retained passages
follow candidate-pool order, and the earliest unselected candidates fill any
remaining positions up to $k$. The complete selector protocol appears in the
supplementary material.

\paragraph{Source grounding.}
After $S_k$ is fixed, source grounding extracts demand-linked verbatim spans
from the selected passages. Candidate annotations are proposed by
\begin{equation}
\widehat W_q=\operatorname{Extract}(q,R_q,S_k).
\end{equation}
The joint grounding call uses $I_j$ and $O_j$ to populate the
\texttt{consumes} and \texttt{produces} bindings. Evidence lineage preserves
demand identity across stages without sequential query rewriting.
The extraction prompt is:
\begin{promptbox}{Source Grounding Prompt}
For each supported Evidence Demand, return its source passage and the verbatim
span that establishes it. Return no record for unsupported demands.
\end{promptbox}
The proposed annotations are filtered by
\begin{equation}
W_q=\{w\in\widehat W_q:\operatorname{Valid}_q(w)\}.
\end{equation}
Validity first anchors the proposal to a selected source: $d_w\in S_k$ and
$s_w\sqsubseteq_{\rm norm}\operatorname{text}(d_w)$. Here
$\sqsubseteq_{\rm norm}$ is substring containment after whitespace and case
normalization. The schema check $V_{\Gamma}(a_w)=1$ verifies the typed
annotation. Provenance continuity further requires
$\exists\ell\,(i_w,\ell)\in\mathcal A_q(d_w)$. These conditions jointly define
$\operatorname{Valid}_q(w)$.

Duplicate passage--span pairs are removed before presentation. The reader
receives the retained annotations $W_q$ alongside the selected passages $S_k$.
A lineage remains open when $W_q$ contains no valid annotation for its demand.
The complete grounding protocol appears in the supplementary material.

\section{Experiments}

We evaluate \method{} through four research questions:
\begin{itemize}
\item \textbf{RQ1.} How does \method{} compare with leading GraphRAG methods
on multi-hop QA benchmarks?
\item \textbf{RQ2.} How do the components of \method{} contribute to
retrieval and answer quality?
\item \textbf{RQ3.} How do evidence-set construction and source grounding account for
\method{}'s retrieval and QA gains?
\item \textbf{RQ4.} What is the computational cost of \method{}?
\end{itemize}

\subsection{Experimental Setup}

\paragraph{Datasets.}
We evaluate on three multi-hop benchmarks: HotpotQA \citep{yang2018hotpotqa}, 2WikiMultiHopQA \citep{ho2020twowiki}, and MuSiQue \citep{trivedi2022musique}. Following HippoRAG~2 \citep{gutierrez2025hipporag2}, we use 1{,}000 queries from each dataset and its corresponding retrieval corpus.

\paragraph{Baselines.}
The baselines are grouped as follows:
\textbf{(1) LLM-based retrieval:} \textbf{NV-Embed-v2}
\citep{lee2025nvembed}.
\textbf{(2) Conventional RAG:} \textbf{BM25}
\citep{robertson2009bm25} and \textbf{RAPTOR}
\citep{sarthi2024raptor}.
\textbf{(3) GraphRAG:} \textbf{HippoRAG~2}
\citep{gutierrez2025hipporag2}, \textbf{PropRAG}
\citep{wang2025proprag}, \textbf{HGRAG} \citep{wang2026hgrag}, and
\textbf{G-reasoner} \citep{luo2026greasoner}.
\textbf{(4) Evidence selection methods:} \textbf{ETS-7B}
\citep{sun2025ets}, \textbf{Relink} \citep{huang2026relink}, and
\textbf{NeocorRAG} \citep{peng2026neocorrag}.

\paragraph{Metrics.}
Retrieval is evaluated with supporting-passage R@5, and QA with exact match
(EM) and token-level F1. Source Grounding is evaluated by demand coverage and
GPT-4o demand--span support judgments, with GPT-4o-mini judging the same samples
independently. The full audit protocol is provided in the supplementary material.

\paragraph{Implementation details.}
Qwen3-32B \citep{qwen2025qwen3} constructs an OpenIE graph with passage and
entity nodes connected by extracted relations, passage--entity edges, and
entity-similarity edges. At query time, it runs the three \method{} stages.
$\operatorname{Retrieve}_{\mathcal G}$ uses HippoRAG~2 PPR with a damping factor
of 0.5, and the matched baseline uses the same graph index and retrieval
settings. NV-Embed-v2 \citep{lee2025nvembed} supplies dense embeddings, and
GPT-4o-mini \citep{openai2024gpt4omini} serves as the QA reader. All generative
calls use temperature 0. Runnable baselines use the best settings reported in
their papers. All rerun methods use the same corpus, query split, and
GPT-4o-mini reader. ETS-7B and Relink use the published results obtained with
their native readers. Full prompts, retrieval settings, candidate budgets, and
baseline configurations are provided in the supplementary material.

\subsection{Main Results (RQ1)}

\begin{table*}[t]
\centering
\small
\setlength{\tabcolsep}{1.5pt}
\begin{tabular}{l*{12}{r}}
\toprule
& \multicolumn{3}{c}{HotpotQA}
& \multicolumn{3}{c}{2WikiMultiHopQA}
& \multicolumn{3}{c}{MuSiQue}
& \multicolumn{3}{c}{Average} \\
\cmidrule(lr){2-4}\cmidrule(lr){5-7}\cmidrule(lr){8-10}\cmidrule(lr){11-13}
Method & R@5 & EM & F1 & R@5 & EM & F1 & R@5 & EM & F1 & R@5 & EM & F1 \\
\midrule
\multicolumn{13}{l}{\emph{LLM-based retrieval}} \\
NV-Embed-v2 (dense) & 93.40 & 59.50 & 73.00 & 75.90 & 55.00 & 60.50
                    & 68.30 & 35.20 & 46.20 & 79.20 & 49.90 & 59.90 \\
\midrule
\multicolumn{13}{l}{\emph{Conventional RAG}} \\
BM25   & 72.90 & 49.40 & 61.00 & 65.80 & 45.10 & 49.00
       & 47.80 & 20.50 & 28.00 & 62.20 & 38.30 & 46.00 \\
RAPTOR & 90.90 & 59.00 & 72.00 & 71.70 & 49.60 & 54.40
       & 65.20 & 31.50 & 41.50 & 75.90 & 46.70 & 56.00 \\
\midrule
\multicolumn{13}{l}{\emph{GraphRAG}} \\
HippoRAG~2 & 95.30 & 58.40 & 72.60 & 89.78 & 60.10 & 69.18
           & 73.04 & 35.70 & 49.92 & 86.04 & 51.40 & 63.90 \\
HGRAG      & 95.50 & 58.70 & 73.90 & 81.00 & 56.50 & 63.00
           & 70.10 & 36.80 & 48.10 & 82.20 & 50.67 & 61.67 \\
PropRAG    & 96.25 & 58.10 & 73.97 & 92.95 & 61.30 & 69.01
           & 73.21 & 35.00 & 49.69 & 87.47 & 51.47 & 64.22 \\
G-reasoner & 97.00 & 59.00 & 74.00 & 93.70 & \textbf{70.00} & \textbf{78.10}
           & 74.90 & 37.50 & 51.50 & 88.53 & 55.50 & 67.87 \\
\midrule
\multicolumn{13}{l}{\emph{Evidence selection methods}} \\
ETS-7B\rlap{$^{\dagger}$}
          & -- & 60.50 & 57.70 & -- & 60.50 & 54.80
          & -- & 40.00 & 40.80 & -- & 53.70 & 51.10 \\
Relink\rlap{$^{\dagger}$}
          & -- & 58.50 & 72.20 & -- & 55.80 & 70.40
          & -- & 30.40 & 41.30 & -- & 48.23 & 61.30 \\
NeocorRAG & 96.00 & 58.40 & 73.67 & 90.90 & 60.80 & 68.50
          & 73.10 & 37.10 & 48.10 & 86.67 & 52.10 & 63.42 \\
\midrule
\method{} & \textbf{98.15} & \textbf{62.76} & \textbf{75.27}
              & \textbf{94.03} & 69.20 & 77.86
              & \textbf{76.53} & \textbf{42.70} & \textbf{55.76}
              & \textbf{89.57} & \textbf{58.22} & \textbf{69.63} \\
\bottomrule
\end{tabular}
\caption{Main results on three multi-hop QA benchmarks. $^{\dagger}$ indicates
published answer metrics obtained with each method's native reader. Average
denotes the macro-average across the three datasets.}
\label{tab:main-results}
\end{table*}

\begin{figure}[!b]
\centering
\includegraphics[width=\columnwidth]{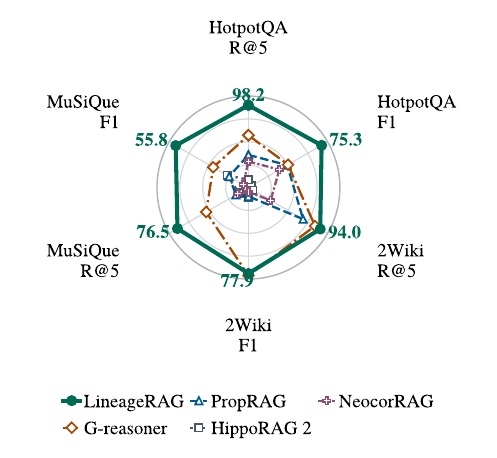}
\caption{Relative R@5 and F1 profiles across three multi-hop QA benchmarks.
Each axis is scaled to its observed range. Values mark \method{} results.}
\label{fig:main-results-radar}
\end{figure}

As shown in Table~\ref{tab:main-results}, \method{} achieves the highest
macro-averaged retrieval and end-to-end answer scores across the three
benchmarks: 89.57 R@5, 58.22 EM, and 69.63 F1.
Figure~\ref{fig:main-results-radar} compares \method{} with leading GraphRAG
methods on R@5 and F1 across the three datasets. Each axis is scaled to its
observed range.

\paragraph{Compared with LLM-based retrieval and conventional RAG.}
The largest margins occur on 2WikiMultiHopQA. \method{} surpasses
NV-Embed-v2 by 18.13 R@5, 14.20 EM, and 17.36 F1 points. The corresponding
gains over RAPTOR are 22.33, 19.60, and 23.46 points. The retrieval advantage
is accompanied by substantial gains in both answer metrics, showing that the
additional supporting passages form useful reader contexts.

\paragraph{Compared with leading GraphRAG baselines.}
\method{} ranks first in seven of the nine dataset--metric cells among
GraphRAG systems and achieves the highest R@5 on every dataset. The
macro-average gains over leading GraphRAG baselines are 3.51 R@5, 5.96 EM,
and 5.22 F1 points. On 2WikiMultiHopQA, \method{} leads in R@5, while the best
GraphRAG EM and F1 scores are 0.80 and 0.24 points higher.

Relative to HippoRAG~2, the R@5 gains are 2.85
points on HotpotQA, 4.25 on 2WikiMultiHopQA, and 3.49 on MuSiQue. The
corresponding end-to-end F1 gains are 2.67, 8.68, and 5.84 points. The largest
retrieval gain and answer gain therefore coincide on 2WikiMultiHopQA, where the
selected and grounded two-hop evidence has the strongest effect. HotpotQA is
closer to retrieval saturation: \method{} exceeds the strongest GraphRAG
R@5 by 1.15 points while improving the strongest GraphRAG EM by 3.76 points.
This separation shows that source-grounded reader input remains consequential
when document recall is already high.

\paragraph{Compared with evidence-selection baselines.}
Among evidence-selection baselines under the aligned reader protocol,
NeocorRAG is the strongest directly comparable system. \method{} raises
average R@5 from 86.67 to 89.57, average EM from 52.10 to 58.22, and average F1
from 63.42 to 69.63. These gains are 2.90, 6.12, and 6.21 points,
respectively. The answer gains exceed the retrieval gain, indicating that
grounding makes the selected evidence more useful to the QA reader.
ETS-7B and Relink are included as published native-reader results, with macro
EM/F1 scores of 53.7/51.1 and 48.23/61.30, respectively.

\begin{samepage}
\subsection{Ablation Study (RQ2)}

We ablate the three \method{} stages on 1{,}000 MuSiQue queries. Without
Evidence-Demand Induction, the original question serves as a single Evidence
Demand. Without Lineage Expansion, the induced demands remain, but candidates
are retrieved with the whole question and enter evidence selection without
demand--passage provenance. Without Source Grounding, the selected passages
and their order remain fixed, while grounded spans are removed from the reader
input. All other stages and settings remain unchanged. Table~\ref{tab:ablation}
reports R@5, EM, and F1.
\end{samepage}
\FloatBarrier

\begin{table}[t]
\centering
\small
\setlength{\tabcolsep}{3pt}
\begin{tabular}{@{}p{0.60\columnwidth}rrr@{}}
\toprule
Variant & R@5 & EM & F1 \\
\midrule
Full \method{} & \textbf{76.53} & \textbf{42.70} & \textbf{55.76} \\
w/o Evidence-Demand Induction & 75.61 & 40.86 & 54.56 \\
w/o Lineage Expansion & 74.38 & 41.12 & 54.48 \\
w/o Source Grounding & 76.53 & 36.50 & 50.11 \\
\bottomrule
\end{tabular}
\caption{Component ablations on MuSiQue.}
\label{tab:ablation}
\end{table}

Removing Evidence-Demand Induction lowers R@5, EM, and F1 by 0.92, 1.84, and
1.20 points. Removing Lineage Expansion produces the largest retrieval loss,
reducing R@5 by 2.15 points, with corresponding decreases of 1.58 EM and 1.28
F1. Source Grounding leaves R@5 unchanged but contributes 6.20 EM and 5.65 F1
points. The first two components improve evidence discovery and selection,
while Source Grounding primarily improves how the reader uses the selected
evidence.

\FloatBarrier
\begin{samepage}
\subsection{Mechanism Analysis (RQ3)}

\method{} constructs a complementary evidence set from demand--passage
provenance and completes supported lineages through Source Grounding. On
MuSiQue, RQ3 examines how evidence-set construction improves retrieval and
whether the grounded spans support their demands and improve answer generation.
\end{samepage}
\FloatBarrier

\paragraph{Evidence-set construction.}
\textbf{\emph{Does set-wise selection alone explain the retrieval gain?}}
Table~\ref{tab:blind-selector} asks whether \method{}'s set-wise selector can
recover the retrieval gain when applied to HippoRAG~2 candidates without
demand--passage provenance. HippoRAG~2 produces 73.04 R@5 from a single graph ranking. Applying
the same set-wise selector to its candidate pool raises R@5 to 74.34, showing
that joint selection contributes 1.30 points without evidence lineage. This
selector can only choose among candidates exposed by the original ranking, and
its input contains no demand--passage provenance. \method{} reaches 76.53 by
organizing candidate discovery around Evidence Demands and retaining their
passage associations during selection. Its 3.49-point gain therefore includes
1.30 points from set-wise selection and a further 2.19 points from
evidence-lineage construction. The comparison shows that reranking improves
evidence coverage, but the larger gain occurs when candidate discovery and
selection preserve the demands associated with each passage. Tables~\ref{tab:mechanism-comparison}
and~\ref{tab:selector-signals} next separate candidate exposure from provenance
within this additional gain.

\begin{table}[t]
\centering
\small
\setlength{\tabcolsep}{4pt}
\begin{tabular}{@{}p{0.68\columnwidth}rr@{}}
\toprule
Evidence-set construction & R@5 & $\Delta$ \\
\midrule
HippoRAG~2 & 73.04 & -- \\
+ Set-wise selector & 74.34 & +1.30 \\
\method{} evidence-set construction & \textbf{76.53} & \textbf{+3.49} \\
\bottomrule
\end{tabular}
\caption{Attribution of evidence-set construction gains on MuSiQue.}
\label{tab:blind-selector}
\end{table}

\textbf{\emph{Do question decomposition and evidence-set selection together
explain the retrieval gain?}}
Table~\ref{tab:mechanism-comparison} compares \method{} with QD, SETR, and their
combination. QD follows the question-decomposition pipeline of
\citet{ammann2025decomposition}. SETR \citep{lee2025setr} performs
explicit-requirement evidence-set selection. Pool oracle measures support
available before selection. Selected reports R@5 after retaining five passages
and Loss is their difference. Best values in each column, including ties, are
bold. QD and SETR select 75.30 and 73.58 R@5. Their
combination raises selected R@5 to 75.47 but remains 1.06 points below
\method{}. QD + SETR and \method{} share a pool-oracle R@5 of 84.26, so lower
aggregate support coverage does not explain the gap. Their selection losses are
8.79 and 7.73. The composition therefore retains less available support under
the five-passage budget. Table~\ref{tab:selector-signals} tests this mechanism by
varying candidate-specific demand--passage provenance.

\begin{table}[t]
\centering
\small
\setlength{\tabcolsep}{3pt}
\begin{tabular}{@{}lrrr@{}}
\toprule
Pipeline & Pool oracle & Selected & Loss \\
\midrule
Question decomposition & 84.02 & 75.30 & 8.72 \\
SETR & 84.24 & 73.58 & 10.66 \\
QD + SETR & \textbf{84.26} & 75.47 & 8.79 \\
\method{} & \textbf{84.26} & \textbf{76.53} & \textbf{7.73} \\
\bottomrule
\end{tabular}
\caption{Candidate availability and five-passage selection on MuSiQue.}
\label{tab:mechanism-comparison}
\end{table}

\begin{table}[!ht]
\centering
\small
\setlength{\tabcolsep}{4pt}
\begin{tabular}{@{}p{0.68\columnwidth}rr@{}}
\toprule
Demand--passage provenance & R@5 & $\Delta$ \\
\midrule
Correct & \textbf{76.53} & -- \\
Removed & 73.58 & -2.95 \\
Equal-length neutral replacement & 73.50 & -3.03 \\
Shuffled across passages & 68.42 & -8.11 \\
\bottomrule
\end{tabular}
\caption{Effect of demand--passage provenance on evidence selection.}
\label{tab:selector-signals}
\end{table}

\textbf{\emph{Does evidence selection depend on correct demand--passage
provenance?}}
All conditions in Table~\ref{tab:selector-signals} use the same candidate pool,
selector, and five-passage output. They differ only in the provenance shown for
each candidate. Removing provenance lowers R@5 from
76.53 to 73.58. Equal-length neutral text gives 73.50, only 0.08 points lower,
so extra prompt content does not explain the gain. Shuffling valid provenance
across passages preserves its volume but breaks the candidate-specific
correspondence. R@5 falls to 68.42, which is 8.11 points below correct
provenance and 5.16 points below removing it. This ordering is diagnostic.
Missing provenance removes demand-coverage information, while incorrect
provenance actively directs its five positions toward the wrong passages.
Correct provenance lets the selector allocate those positions across distinct
Evidence Demands and retain complementary support. Read with
Table~\ref{tab:mechanism-comparison}, these controls attribute the selection-side
benefit of evidence lineage to preserving candidate-specific demand--passage
correspondence between Demand-Conditioned Lineage Expansion and set-wise
selection.

\paragraph{Source Grounding.}
\textbf{\emph{How often does Source Grounding complete an evidence lineage?}}
Table~\ref{tab:grounding-validation} measures whether the lineage established
for each Evidence Demand reaches a verbatim span in the selected source.
Source Grounding completes 1{,}865 of 2{,}182 induced lineages. All lineages are
completed for 734 of 1{,}000 queries, and 976 queries contain at least one
completed lineage. Of the 266 queries without full completion, 242 contain
both completed and open lineages, while 24 contain no completed lineage.
Open lineages are thus concentrated within partially completed queries.
Demand-level analysis identifies requirements still ungrounded after
evidence-set construction.

\textbf{\emph{Do the grounded spans support their linked Evidence Demands?}}
Table~\ref{tab:grounding-validation} evaluates whether each grounded endpoint
establishes its linked demand. Of the 500 audited spans, 420 provide full
support and 50 provide partial support. The partial cases reach relevant source
evidence but do not establish the complete fact specified by the demand. Among
the remaining 30 records, 20 do not establish the required fact, while 10
attach a span to the wrong demand. These errors distinguish insufficient source
evidence from an incorrect demand--span connection. The counts expose span
adequacy and correspondence as separate factors. GPT-4o-mini matches GPT-4o on
481 of 500 binary support decisions (96.2\%), indicating that the distinction
between supporting and insufficient evidence is stable across the two judges.

\begin{table}[t]
\centering
\small
\setlength{\tabcolsep}{3pt}
\begin{tabular}{@{}lr@{}}
\toprule
Measure & Result \\
\midrule
\multicolumn{2}{l}{\emph{Evidence-lineage completion}} \\
Grounded Evidence Demands & 1{,}865 / 2{,}182 demands \\
Queries with all demands grounded & 734 / 1{,}000 queries \\
Queries with any demand grounded & 976 / 1{,}000 queries \\
\midrule
\multicolumn{2}{l}{\emph{Demand--span support}} \\
Fully supporting spans & 420 / 500 audited spans \\
Full or partial support & 470 / 500 audited spans \\
Wrong-demand links & 10 / 500 audited spans \\
\bottomrule
\end{tabular}
\caption{Evidence-lineage completion on 1{,}000 MuSiQue queries and
demand--span support on 500 audited spans.}
\label{tab:grounding-validation}
\end{table}

\begin{table}[t]
\centering
\small
\setlength{\tabcolsep}{5pt}
\begin{tabular}{@{}p{0.66\columnwidth}rr@{}}
\toprule
Reader evidence & EM & F1 \\
\midrule
Selected passages & 36.50 & 50.11 \\
Selected passages with grounded spans & \textbf{42.70} & \textbf{55.76} \\
Gain & \textbf{+6.20} & \textbf{+5.65} \\
\bottomrule
\end{tabular}
\caption{Contribution of Source Grounding to answer generation on MuSiQue.}
\label{tab:grounded-reader-control}
\end{table}

\begin{table}[t]
\centering
\small
\setlength{\tabcolsep}{3pt}
\begin{tabular}{@{}lrrr@{}}
\toprule
Method & Offline & Query-time & Total \\
& (M) & (M / 1{,}000 queries) & (M) \\
\midrule
\method{} & 13.49 & 1.51 & 15.00 \\
HippoRAG~2 & 13.49 & 0.63 & 14.12 \\
PropRAG & 18.84 & 0.00 & 18.84 \\
NeocorRAG & 12.84 & 6.94 & 19.78 \\
\bottomrule
\end{tabular}
\caption{Average model-token consumption across the three benchmarks
(millions of tokens).}
\label{tab:aux-token-cost}
\end{table}

\textbf{\emph{Does Source Grounding improve answer generation?}}
\mbox{Table~\ref{tab:grounded-reader-control}} fixes the selected passages and
their order, changing only whether grounded spans are provided. Adding them
raises EM from 36.50 to 42.70 and F1 from 50.11 to 55.76. Because retrieval is
identical, the spans introduce no new sources and leave the passage set and order
unchanged, isolating their effect on reader-side evidence use. In the
passage-only condition, the reader must locate support within the passages and
infer which demand each passage resolves before combining facts. Grounded spans
make both decisions explicit through demand-indexed evidence cues, allowing the
reader to compose facts whose locations and roles are already identified. The
gains in both EM and F1 indicate better exact-answer recovery and broader
answer-token coverage. Source Grounding therefore improves how the reader uses
selected evidence while leaving evidence availability unchanged.

The mechanism analysis follows evidence lineage from evidence-set construction
to answer generation. Tables~\ref{tab:blind-selector}--
\ref{tab:selector-signals} show that preserving demand--passage provenance
improves the selected evidence set. Table~\ref{tab:grounded-reader-control}
shows that carrying the same demand-to-evidence correspondence into source spans improves how
the reader uses that evidence. The two analyses locate the effect at successive
stages: demand--passage correspondence guides which evidence is selected, and
demand--span correspondence guides how selected evidence is presented to the
reader.

\subsection{Computational Cost (RQ4)}

Table~\ref{tab:aux-token-cost} separates one-time graph construction from
query-time model use. \method{} and HippoRAG~2 share the same graph index, whose
construction consumes 13.49M tokens. HippoRAG~2 follows one whole-query retrieval path. \method{} additionally
induces demands, retrieves for each demand, selects from a
provenance-augmented pool, and extracts grounded spans for the reader. These
stages preserve demand identity from candidate discovery through source grounding and increase
query-time token consumption from 0.63M to 1.51M tokens per 1{,}000 queries, or 880 additional
tokens per query. The total for one graph construction and 1{,}000 queries rises
from 14.12M to 15.00M. Because graph construction is shared, the total
difference comes from these query-time lineage operations.

This 2.40-fold increase in query-time token consumption accompanies gains of 3.53 R@5, 6.82 EM, and
5.73 F1 over HippoRAG~2. The ablations attribute 2.15 R@5 points to Lineage
Expansion and 6.20 EM and 5.65 F1 points to Source Grounding. These gains locate
the return from the additional computation in both evidence construction and
answer generation. PropRAG uses 18.84M tokens offline and none online, while
NeocorRAG uses 6.94M online.

\FloatBarrier

\section{Conclusion}

In this work, we propose \method{} for multi-hop GraphRAG to address the missing
connection between evidence discovery and source grounding. \method{} induces
Evidence Demands and expands each through graph retrieval while preserving
demand--passage provenance. It then uses this provenance to construct a
complementary evidence set and ground supported demands in source text.
Experiments on three multi-hop QA benchmarks show improved retrieval and answer
quality over leading GraphRAG baselines and demonstrate the value of evidence lineage
for source-grounded multi-hop reasoning.

\bibliography{references}

\end{document}